\documentstyle[11pt,epsfig]{article}
 
\newcommand{\beq}{\begin{equation}}
\newcommand{\eeq}{\end{equation}}
\newcommand{\lae}{\begin{array}{c}\,\sim\vspace{-21pt}\\< \end{array}}
\newcommand{\gae}{\begin{array}{c}\,\sim\vspace{-21pt}\\> \end{array}}

\begin{document}
\begin{titlepage}
\def\thepage {}        
 
\title{A Heavy Top Quark From Flavor-Universal Colorons}

\author{Marko B. Popovic and Elizabeth H. Simmons
\thanks{e-mail addresses: markopop@buphy.bu.edu,
  simmons@bu.edu.}\\
Department of Physics, Boston University, \\
590 Commonwealth Ave., Boston MA  02215}
 
\date{\today}

\maketitle
 
\bigskip
\begin{picture}(0,0)(0,0)
\put(295,250){BUHEP-98-10}
\put(295,235){hep-ph/9806287}
\end{picture}
\vspace{24pt}
 
\begin{abstract}

  Ordinary technicolor and extended technicolor cannot produce the heavy
  top quark unaided.  We demonstrate that a flavor-universal extension
  of the color interactions combined with an extended hypercharge sector
  that singles out the third generation can provide the necessary
  assistance.  We discuss current experimental constraints and suggest
  how collider experiments can search for the predicted new heavy gauge
  bosons.

\pagestyle{empty}
\end{abstract}
\end{titlepage}
 
\section{Introduction}
\label{sec:intro}
\setcounter{equation}{0}
 
Generating mass through strong gauge dynamics is a challenge.  While a
technicolor \cite{tc} gauge sector can provide appropriate masses for
the electroweak gauge bosons by breaking the chiral symmetries of
technicolored fermions, explaining the masses and mixings of the quarks
and leptons has proven more difficult.  Extended technicolor (ETC)
models \cite{etc} postulate an enlarged gauge group coupling the quarks
and leptons to the technifermion condensate, enabling them to acquire
mass.  The simplest models of this type tend to produce large
flavor-changing neutral currents \cite{etc} and, if the heavy top quark
mass is generated by ETC interactions, excessive weak isospin violation
\cite{ta} and contributions to $R_b$ \cite{css}.  Substantially raising
the scale at which extended technicolor breaks to its technicolor
subgroup can alleviate some of these problems -- but renders the model
incapable of naturally producing quark masses larger than a few GeV.

Given the large value of the top quark's mass ($m_t \approx 175$ GeV
\cite{pdg}) and the sizable splitting between the masses of the top and
bottom quarks, it is natural to wonder whether $m_t$ has a different
origin than the masses of the other quarks and leptons.  A variety of
dynamical models that exploit this idea have been proposed.  Key
examples are the dynamical models of `top-mode' mass generation in which
top quark self-interactions drive all of electroweak symmetry breaking
\cite{tmode}.  Related to those are the topcolor \cite{topcolor} and
topcolor-assisted technicolor \cite{topassist} models
\cite{topastmod,bbhk} in which the top quark feels different color and
hypercharge interactions than other quarks; as a consequence, a top
quark condensate enhances the top quark's mass.  Finally, there are the
non-commuting ETC scenarios where the top quark has weak and extended
technicolor interactions different from those of other quarks
\cite{ncETC}.  The conclusion of these investigations has been that new
dynamics peculiar to the top quark can certainly create a large top
quark mass.  It may even possible to do so while creating a model that
accords reasonably well with electroweak precision data.

In this paper, we discuss a variant class of models of dynamical top
quark mass generation in which the large mass comes from top-specific
gauge interactions.  What sets these theories apart is that the top
quark differs from the other quarks only in its hypercharge
interactions.  The extended color interactions are flavor-universal,
just as in the coloron model of \cite{coloron}; the weak interactions
display ordinary Cabibbo universality.

After introducing the class of models in section 2 and showing, in
section 3, that the low-energy dynamics admit the possibility of top
quark condensation and a large top quark mass, we focus on experimental
constraints.  Section 4 discusses the phenomenology of the low-energy
effective theory, while section 5 explores the possibility of direct
searches for the additional massive gauge bosons in the theory.

We note that the physics discussed here must be part of some larger
(e.g. ETC) structure at high energies which will create the masses and
mixings of the light fermions and produce condensates that break our
extended gauge symmetries to their standard model subgroups.  However,
we focus on exploring the dominant effects of the new physics that
produces the top quark mass.  A discussion of higher-scale operators
that break all fermion chiral symmetries, account for generational
mixing and produce relevant symmetry-breaking condensates may be found
in \cite{KDL}.

\section{The Class of Models}
\label{sec:model}
\setcounter{equation}{0}

Our models have a gauge structure like that of the original
topcolor-assisted technicolor models \cite{topassist}.  Far above the
electroweak scale, the gauge group is
\beq
SU(N)_{TC} \times SU(3)_1 \times SU(3)_2 \times SU(2)_W \times U(1)_1
\times U(1)_2\ \ .
\label{origgrp}
\eeq
with coupling constants $g_{N}$, $g_{3_{(1)}}$, $g_{3_{(2)}}$, $g_2$,
$g_{1_{(1)}}$, and $g_{1_{(2)}}$ respectively. We take the first $SU(3)$
and $U(1)$ groups to have the stronger couplings: $g_{3_{(1)}} >
g_{3_{(2)}}$ and $g_{1_{(1)}} > g_{1_{(2)}}$.  The group $SU(N)_{TC}$ is
the technicolor gauge group.

At an energy scale $\Lambda$, a condensate $\langle\phi\rangle$
transforming under the initial symmetry group as $(1, 3, \bar{3}, 1, p,
-p)$ breaks the color sector ($SU(3)_1 \times SU(3)_2$) to its diagonal
subgroup ($SU(3)_{C}$) and similarly breaks the hypercharge groups in
the pattern ($U(1)_1\times U(1)_2 \rightarrow U(1)_{Y}$).  The gauge
symmetry is reduced to that of the standard model plus the unbroken
technicolor group:
\beq
SU(N)_{TC} \times SU(3)_{C} \times SU(2)_{W} \times U(1)_{Y}\ \ .
\eeq
At the weak scale, $\Lambda_{EW}<\Lambda$, the technicolor force becomes
strong enough to break the chiral symmetries of a set of technifermions
and cause electroweak symmetry breaking $SU(2)_W \times U(1)_Y \to
U(1)_{EM}$.  Thus the low-energy gauge boson spectrum includes the
massless photon and gluons, the massive $W$'s and $Z$, and two
additional kinds of massive states: an octet of colorons and a single
$Z'$.

At low energies, the mass eigenstate fields in the color sector
(colorons $C^a$ and gluons $G^a$) are related to the original $SU(3)_1
\times SU(3)_2$ gauge fields (denoted $X_{(n)}$) via \cite{hlpk}:
\beq
G^a={g_{3_{(2)}} X_{(1)}^a + g_{3_{(1)}} X_{(2)}^a \over \ 
\sqrt{g_{3_{(1)}}^2 +
g_{3_{(2)}}^2}} \qquad  \qquad  C^a={g_{3_{(1)}} X_{(1)}^a - 
g_{3_{(2)}} X_{(2)}^a \over 
\ \sqrt{g_{3_{(1)}}^2 + g_{3_{(2)}}^2}} \ \ .
\eeq
Similar relations hold in the hypercharge sector. The familiar QCD and
hypercharge gauge coupling constants are related to the high energy
couplings by
\beq
{1\over g_{3}^{2}}={1\over g_{3_{(1)}}^{2}}+ {1\over g_{3_{(2)}}^{2}}
\qquad  \qquad {1\over g_{1}^{2}}={1\over g_{1_{(1)}}^{2}}+ {1\over 
g_{1_{(2)}}^{2}}
\label{ggggs}
\eeq   
and their respective fine-structure constants are $\alpha_Y \equiv
g_1^2/4\pi$ and $\alpha_s \equiv g_3^2/4\pi$.  The tree level masses of
the colorons and $Z'$ are
\beq  
M_{C}= \langle\phi\rangle 
\sqrt{g_{3_{(1)}}^2 + g_{3_{(2)}}^2}  \qquad  \qquad 
M_{Z'}=\langle\phi\rangle \, |p| \, 
\sqrt{g_{1_{(1)}}^2 + g_{1_{(2)}}^2}\ \ .
\eeq
Note the dependence of the $Z'$ mass on the $U(1)$ charges of the
condensate $\langle\phi\rangle$.

The gauge transformation properties of the quarks and leptons, which are
summarized in Table 1, are significantly different from those in
topcolor-assisted technicolor \cite{topassist}.  In the color sector,
all quarks transform only under the stronger $SU(3)_{1}$ group, as in
the flavor-universal coloron model \cite{coloron}.  In the hypercharge
sector only the third family of fermions transforms under the stronger
$U(1)_{1}$ and the first two families transform under the weaker
$U(1)_{2}$ (all of them with standard model hypercharge assignments).
All of the quarks and leptons have the same weak charge assignments as
in the standard model.  Each generation of ordinary fermions forms an
anomaly-free representation of the gauge group (\ref{origgrp}).

As we shall explore in more detail, this set of gauge charge assignments
for the fermions still allows natural dynamical generation of a large
mass for the top quark (and only the top quark).  Yet it leads to a
phenomenology differing from that of topcolor-assisted technicolor
\cite{topassist,bbhk}.

\begin{table}[bhtp]
\begin{center}
\begin{tabular}{|c||c|c|c|c|c|c||}\hline\hline
 & $SU(N)_{TC}$ & $SU(3)_1$ & $SU(3)_2$ & $SU(2)$ & $U(1)_1$ & $U(1)_2$ 
\\ \hline \hline
I & 1 & SM & 1 & SM & 0 & SM \\
II& 1 & SM & 1 & SM & 0 & SM \\
III & 1 & SM & 1 & SM & SM & 0 \\
\hline\hline
\end{tabular}
\end{center}
\vspace{-.5cm}
\caption{\small Quark and lepton gauge charge assignments 
  for generations I, II and III.  An entry of `SM' indicates that 
  the particles carry the same charges under the given group as they 
  would under the standard model group of the same rank.}
\label{Pred}
\end{table}

\section{Low energy effective theory}
\label{sec:loweneft}
\setcounter{equation}{0}

Below the symmetry-breaking scale, $\Lambda$, for the extended color and
hypercharge sectors, the interactions among quarks and leptons that
arise from exchange of the massive colorons and $Z'$ are
well-approximated by effective four-fermion interactions
\beq
{\cal L}_{C}= - {2 \pi \kappa_{3} \over M_{C}^2} 
\left(\bar{q} {\gamma}^{\mu} {{{\lambda}^{a}} \over2} q\right) 
\left(\bar{q} {\gamma}_{\mu} {{{\lambda}_{a}} \over2} q\right) 
\label{effcoleq}
\eeq
\begin{eqnarray}
\nonumber\
{\cal L}_{Z'}&=& -{2 \pi
\over {M_{Z'} ^2}} {\alpha_{Y}^2 \over \kappa_{1}}
\left( \bar{f}_{\!\!\!\scriptscriptstyle{\atop{I \atop {II}}}} 
{\gamma}^{\mu}{Y \over 2} \;
f_{\!\!\!\scriptscriptstyle{\atop{I\atop{II}}}}\right)
\left( \bar{f}_{\!\!\!\scriptscriptstyle{\atop{I \atop {II}}}} 
{\gamma}_{\mu}{Y \over 2} \;
f_{\!\!\!\scriptscriptstyle{\atop{I\atop{II}}}}\right) \\
\nonumber\
&&-{2 \pi \kappa_{1} \over {M_{Z'} ^2}}
\left(\bar{f}_{\scriptscriptstyle III}\; 
{\gamma}^{\mu}{Y \over 2}\;  f_{\scriptscriptstyle III}\right)
\left(\bar{f}_{\scriptscriptstyle III}\; 
{\gamma}_{\mu}{Y \over 2}\;  f_{\scriptscriptstyle III}\right)\\
&&+{4 \pi \alpha_{Y} \over {M_{Z'} ^2}} 
\left( \bar{f}_{\!\!\!\scriptscriptstyle{\atop{I \atop {II}}}} 
{\gamma}^{\mu}{Y \over 2} \;
f_{\!\!\!\scriptscriptstyle{\atop{I\atop{II}}}}\right)
\left(\bar{f}_{\scriptscriptstyle III}\;
{\gamma}_{\mu}{Y \over 2}\; f_{\scriptscriptstyle III}\right)
\label{effzpeq}
\end{eqnarray}
where $q$ is any quark, $f$ is a quark or lepton whose subscript
indicates its generation, the $\lambda^{a}$ are the octet of Gell-Mann
matrices, and Y is the standard model hypercharge generator\footnote{We
  use the convention $Q = T_3 + {1\over2} Y$.}.  The coefficients
$\kappa_{1}$ and $\kappa_{3}$ are defined as
\beq 
\kappa_{1}= \alpha_Y \bigg{(}{g_{1_{(1)}} \over 
g_{1_{(2)}}}\bigg{)}^2
\qquad  \qquad \kappa_{3}= \alpha_s \bigg{(}{g_{3_{(1)}} 
\over g_{3_{(2)}}}\bigg{)}^2\ \ .
\label{kapdef}
\eeq
Note that $g_{i_{(1)}} / g_{i_{(2)}} \equiv \cot(\theta_i)$ where
$\theta_i$ is the angle by which the original color (i=3) and
hypercharge(i=1) gauge eigenstates were rotated to form the mass
eigenstates.

The extended gauge interactions are ultimately responsible for the large
mass of the top quark.  The principle contributions to the dynamical
mass come from the four-fermion contact interactions (\ref{effcoleq})
and (\ref{effzpeq}), which we can study using the gap equation in the
Nambu--Jona-Lasinio (NJL) approximation \cite{NJL}.  The dynamical mass
of fermion $f$ is the solution to:
\beq
m_f\; =\;  G_{1} {{m_f M_{Z'}^{2}}\over{8\pi^2}}
\left[1- \left({m_f \over M_{Z'}}\right)^2 
\ln({M_{Z'}^2 \over m_f^2})\right]
\; +\; G_{3}{{3 m_f M_C^{2}}\over{8\pi^2}}
\left[1-\left({m_f \over M_C}\right)^2 \ln({M_C^2 \over m_f^2})\right]
\label{gap}
\eeq
where the coefficients $G_i$ are
\begin{eqnarray}
\nonumber\
G_3 &=& 0 {\rm\ \  for\ leptons}\ \ \ \ \ \ \ \ \  
G_3=4\pi{\kappa_3 \over M_C^2} {\rm\ \  for\ quarks}\\
\nonumber\
G_1 &=& {2\pi \alpha_Y^2 \over{M_{Z'}^2 \kappa_1}} Y^f_L Y^f_R 
{\rm\ \  for\ generations\ I\ and\ II}\\
\nonumber\
G_1 &=& {2\pi \kappa_1 \over{M_{Z'}^2}} Y^f_L Y^f_R 
{\rm\ \ for\ generation\ III}
\end{eqnarray}
and $Y_L^f$ ($Y_R^f$) is the hypercharge of $f_L$ ($f_R$).  In solving
(\ref{gap}), we take the cut-off $\Lambda$ for the gap equation to be of
order the coloron and $Z'$ masses: $\Lambda \sim M_{C} \sim M_{Z'}$;
corrections due to unequal values for the coloron and $Z'$ masses are
small in the region of physical interest.  Applying this to the top
quark, one finds $\langle\,\bar{t}t\,\rangle \neq 0$ if
\beq
\kappa_{3} + {2\over 27}\kappa_{1} \geq {{2 \pi}\over 3} 
\eeq
\label{origress}
More generally, however, we need to include contributions to the gap
equation from gluon and hypercharge boson exchange\footnote{Since the
  $SU(2)_W$ bosons couple only to left-handed fermions, they do not
  contribute here.}; in effect, we are studying a {\em gauged} NJL model
\cite{gaugeNJL}.  As discussed in \cite{gaugeNJLres}, this modifies the
criticality conditions for the $\kappa_i$.

Applying the gauged NJL gap equations to all the standard model
fermions, we seek solutions with non-zero $m_t$ (i.e. formation of a top
condensate $\langle\,\bar{t}t\,\rangle\neq 0$) and no mass for any other
fermion (i.e.  $\langle\,\bar{f}f\,\rangle = 0$ for $f \neq t$).  Such
solutions exist provided that $\kappa_1$ and $\kappa_3$ satisfy a set of
inequalities, of which the following three are the most stringent:
\beq
\kappa_{3} + {2\over 27}\kappa_{1} \geq {{2 \pi}\over 3} -
{4\over3}\alpha_s - {4\over 9}\alpha_Y
\label{kapin1} 
\eeq
\beq         
\kappa_{3} + {2\over 27}{\alpha_{Y}^{2}\over \kappa_{1}} < 
{{2 \pi}\over 3} - {4\over3}\alpha_s - {4\over 9}\alpha_Y
\label{kapin2}
\eeq
\beq
\kappa_{1}\ < 2 \pi - 6 \alpha_Y\ \ .
\label{kapin3}
\eeq
Inequality (\ref{kapin1}) leads to top quark condensation
($\langle\,\bar{t}t\,\rangle\neq 0$).  Note how including the effects of
gluon and hypercharge boson exchange modifies the right-hand-side
expression compared to the original NJL result (\ref{origress}).
Inequality (\ref{kapin2}) implies $\langle\,\bar{c}c\,\rangle= 0$ (i.e.,
no charm quark condensation).  In our class of models, this is a
stronger constraint than the inequality ensuring
$\langle\,\bar{b}b\,\rangle = 0$; in a Top-color I model \cite{bbhk},
the latter would be the relevant constraint.  Inequality (\ref{kapin3})
is related to the lack of $\tau$ condensation; it will be superseded by
other constraints later in our discussion.

As inequalities (\ref{kapin1}) -- (\ref{kapin3}) can be simultaneously
satisfied, our models {\em do} admit the possibility that only the top
quark condenses and receives an enhanced mass.  The values of the
couplings $\kappa_1$ and $\kappa_3$ for which this happens fall within
the `gap triangle'\footnote{Due to the non-linearity of (\ref{kapin2}),
  it is only approximately a triangle.} lying to the right of curve (1),
to the left of curve (2) and below curve (3) in Figure 1 (by analogy
with results for Top-color I models \cite{bbhk}).  Solutions to the
gauged-NJL gap equation \cite{gaugap} for $m_t = 175$ GeV and particular
values of the cut-off $\Lambda \sim M_C \sim M_{Z'}$ lie on curves
parallel to curve (1); a few examples for $\Lambda$ ranging from 0.7 TeV
to 5 TeV are shown and labeled (A) through (D).  Curves like these will
be used in calculating phenomenological limits in the next section.

\begin{figure}
\centerline{\epsfig{file=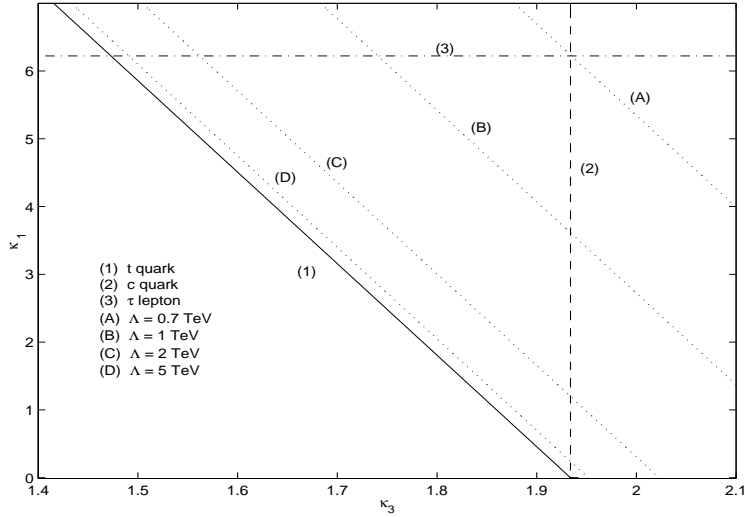, height=10cm,width=7cm,angle=90}}
\caption[one]{\small The gap triangle, bounded by curves (1), (2), and
  (3) is the region within which only the top quark condenses.  Above
   curve (1) $\langle\,\bar t t\,\rangle \neq 0$; to the
  left of curve (2) $\langle\,\bar c c\,\rangle = 0$, and below curve
  (3) $\langle\,\bar \tau \tau\,\rangle = 0$.  Lines (A, B, C, D)
  represent solutions to the gap equation \cite{gaugap} for $m_t = 175$
  assuming $ \Lambda\sim M_{Z'}\sim M_{C}$ has values of (0.7, 1.0, 2.0,
  5.0) TeV.}
\end{figure}

\section{Low-energy constraints}
\label{sec:constraints}
\setcounter{equation}{0}

We now consider how several types of physics constrain the allowed
region of the $\kappa_1-\kappa_3$ plane.  We first look at the $\rho$
parameter and $Z$ decays to tau leptons.  Next, we discuss the
implications of a strong $U(1)_1$ coupling.  Finally, we comment on
flavor-changing neutral currents (FCNC).

Current measurements of the $\rho$ parameter are already sensitive to
the presence of the low energy contact interactions (\ref{effcoleq}) and
(\ref{effzpeq}).  The main contribution to $\Delta\rho_{\ast}$ from the
coloron sector of our model is \cite{IB} single coloron exchange across
the top and bottom quark loops of $W$ and $Z$ vacuum polarization
diagrams. Applying the results of \cite{IB} to our models, we have
\beq
\Delta \rho_{\ast}^{(C)}\approx{{16 {\pi}^2 \alpha_{Y}} \over {3
\sin^2\theta_{W}}} \bigg{(}{f_{t}^2 \over {M_{C} M_{Z}}}\bigg{)}^2
\kappa_{3}
\label{delrhoc}
\eeq  
where $\theta_{W}$ is the weak mixing angle and  $f_{t}$ is the
analog of $f_{\pi}$ for the top-condensate, i.e. (in the  
NJL approximation) \cite{NJL,F} 
\beq
f_{t}^2={3 \over {8\pi^2}} m_{t}^2 \ln\bigg{(}{\Lambda^2 \over
m_{t}^2}\bigg{)}\ \ .
\eeq
In the $Z'$ sector, the main contribution to $\Delta \rho_{\ast}$ arises
from $Z-Z'$ mixing. Adapting the results of \cite{ZZ} to our models, we
have
\beq
\Delta \rho_{\ast}^{(Z')}\approx{\alpha_{Y}\sin^2\theta_{W} \over
\kappa_{1}} {M_{Z}^2 \over M_{Z'}^2}\bigg{[}1-{f_{t}^2 \over
v^2}({\kappa_{1}
\over \alpha_{Y}}+1)\bigg{]}^2\ \ .
\eeq
Requiring $\Delta
\rho_{\ast}=\Delta\rho_{\ast}^{(C)}+\Delta\rho_{\ast}^{(Z')}<0.4\%$
\cite{IB} excludes the region to the right of curve (4) in Figure 2.
This curve connects the points $\Delta \rho_{\ast}=0.4\%$ on the lines
of constant $\Lambda \sim M_{C} \sim M_{Z'}$ mentioned earlier.  Note
how the $\Delta \rho_{\ast}$ constraint narrows the allowed region of
the $\kappa_1-\kappa_3$ plane.

\begin{figure}
\centerline{\epsfig{file=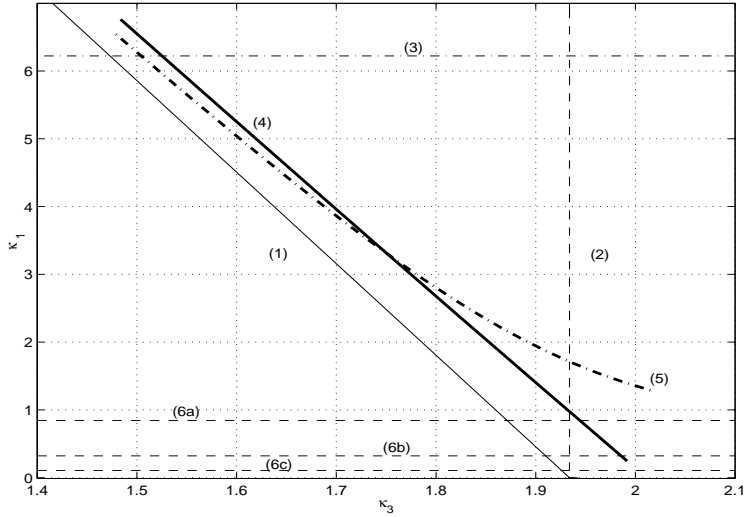, height=10cm,width=7cm,angle=90}}
\caption[two]{\small Low-energy constraints.  Curves (1), (2), (3) 
  outline the `gap triangle' of Figure 1 where only $\langle\,
  \bar{t} t \,\rangle \neq 0$.  The region above curve (4) is 
  excluded by data on $\Delta\rho_{*}$; the region above curve (5) 
  is excluded by data on $Z\rightarrow\tau^{+}\tau^{-}$. Lines 
  (6a-6c) are possible upper bounds on $\kappa_1$ from triviality 
  as in Figure 3.}
\end{figure}

Another constraint comes from the partial decay width of the Z boson to
tau leptons:
\beq
\Gamma(Z\to\tau^{+}\tau^{-})={G_{F} M_{Z}^{3} \over
3\sqrt{2}\pi}\big {[}g_{\tau_{L}}^2 + g_{\tau_{R}}^2 \big {]}
\label{tautau}
\eeq  
where $G_{F}$ is the Fermi constant \cite{pdg} and $g_{\tau_{L}}$
($g_{\tau_{R}}$) is the coupling of $\tau_L$ ($\tau_R$) to the $Z$
boson.  Due to $Z-Z'$ mixing, \cite{ZZ}, the couplings $g_{\tau_{L}}$
and $g_{\tau_{R}}$ in our model are altered from those in the standard
model (i.e. $g_{\tau}\rightarrow g_{\tau}\scriptstyle{(SM)}$ + $ \delta
g_{\tau}$) by
\beq
\delta g_{\tau_{L}}={1\over 2}\delta 
g_{\tau_{R}}=\sin^2\theta_{W}{M_{Z}^2 
\over M_{Z'}^2} \bigg{[}1-{f_{t}^2 \over v^2}({\kappa_{1}\over 
\alpha_{Y}}+1)\bigg{]}\ \ ,
\eeq 
yielding a non-standard prediction for $\Gamma(Z\to\tau^{+}\tau^{-})$.
Including QED corrections to eq. (\ref{tautau}) and requiring our
predicted value to be consistent with the experimental \cite{pdg} value
$\Gamma^{expt}(Z\to\tau^+\tau^-) = 83.67 \pm 0.44 $ MeV at 95\% c.l.
excludes the region to the right of curve\footnote{This curve was
  constructed by the same procedure as curve (4).} (5) in Figure 2.

\begin{figure}
\centerline{\epsfig{file=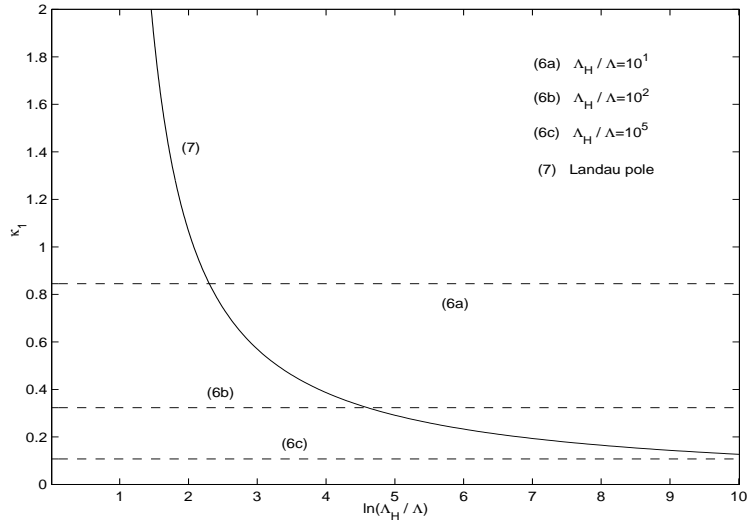, height=10cm,width=7cm,angle=90}}
\caption[thr]{The position of the Landau pole $\Lambda_H$ for 
$U(1)_1$ is shown by curve (7).  Lines (6a-6c) show the upper 
bound on $\kappa_1$ that holds if the Landau pole lies one, 
two or five orders of magnitude above $\Lambda$; these also appear in 
Figure 2.}
\end{figure}

The asymptotic UV behavior of the strongly-coupled $U(1)_1$ yields
another important, albeit elastic, constraint\footnote{We thank
  R.S.~Chivukula for emphasizing the relevance of this constraint.
  Similar considerations apply in any model in which a $U(1)$ gauge
  interaction is used to align the vacuum \cite{rschg}.} on $\kappa_1$.
Combining expressions (\ref{ggggs}) and (\ref{kapdef}) shows that
\beq
{{g_{1}}_{(1)}^2 \over {4\pi}}=\alpha_{Y}+\kappa_1
\eeq
Applying the renormalization group equation to $U(1)_{1}$
\beq
 {{g_{1}}_{(1)}^2 \over {4\pi}}\mid_{\Lambda_{H}}={{{g_{1}}_{(1)}^2 
\over {4\pi}}\mid_{\Lambda}\over {1-({{g_{1}}_{(1)}^2 \over 
{4\pi}})\mid_{\Lambda} {C \over {3\pi}} \ln ({\Lambda_{H}^2 \over 
{A \Lambda^2}})}}
\eeq
(with $A= \exp{({5 \over 3})}$) and considering just the contribution
from the standard model particles (i.e., taking $C={15 \over 4}$) allows
us to estimate the position of the Landau pole for a given low-energy
value of $\kappa_1$.  Our results are summarized in Figure 3.  If the
Landau pole is to lie at least an order of magnitude above the
symmetry-breaking scale, $\Lambda$, then $\kappa_1$ must be of order 1
or smaller.  This defines curve (6a) in Figures 2 and 3. Similarly,
requiring the Landau pole to lie two or five orders of magnitude above
$\Lambda$ produces curves (6b) and (6c) in Figures 2 and 3.

Finally, we turn to flavor-changing neutral currents.  Because the color
sector is flavor-universal, the low-energy effective interactions
(\ref{effcoleq}) cause no flavor-changing neutral currents.  In other
words, the low-energy effective theory now includes not just top-pions
\cite{topassist}, but a complete set of ``q-pions'' strongly coupled to
all flavors of quarks.  To first approximation, the q-pion masses and
couplings are flavor-symmetric and they make no contribution to hadronic
FCNC processes like neutral meson mixing or $b\to s\gamma$.  This is in
contrast to the potentially large (but avoidable) hadronic FCNC
exhibited by Top-color I models \cite{dkom,bbhk}.  The flavor symmetry
among the q-pions will be modified at sub-leading level by non-universal
U(1) effects; this can re-introduce hadronic FCNC at a smaller, less
dangerous rate.

Because the hypercharge interactions (\ref{effzpeq}) distinguish among
generations, they also cause semi-leptonic flavor-changing decays of $B$
and $K$ mesons, which are the same as those in Top-color I models
\cite{bbhk}.  As discussed in ref. \cite{bbhk}, current data on $B_s \to
l^+ l^-$, $B \to X_s l^+ l^-$, $B \to X_s \nu \bar\nu$, and
also\footnote{While this process involves no FCNC, it would be similarly
  affected by the $Z'$ boson.} $\Upsilon(4S) \to l^+ l^-$ set no limits,
but future experiments may be sensitive to the presence of the
additional interactions. For the process $K^+ \to \pi^+ \nu_\tau
\bar\nu_\tau$, ref. \cite{bbhk} found that the ratio of amplitudes was
roughly $\vert A_{new}/A_{SM}\vert \sim 1.5 {\kappa_1/M_{Z'}^2} {\rm
  TeV^2}$, so squaring this and dividing by the number of neutrino
species gives an estimate of the relative branching ratios: $B_{new}(K
\to \pi \nu_\tau \bar\nu_\tau) / B_{SM} (K\to\pi\nu\bar\nu) \sim 0.8
({\kappa_1/M_{Z'}^2})^2 {\rm TeV^4}$.  Subsequently, evidence has been
published for a $K^+ \to \pi^+ \nu \bar\nu$ event that is consistent
with branching ratio $4.2 ^{+9.7}_{-3.5} \times 10^{-10}$
\cite{kpinunu}, as compared with a standard model branching ratio of
order $10^{-10}$.  This process is therefore still able to accommodate a
$Z'$ in the allowed parameter space of our models (i.e., $\kappa_3
\approx 2$ and $\kappa_1 \lae 1$); future data from the E787
Collaboration may provide further constraints.

\section{Direct Searches for the Colorons and $Z'$} 
\label{sec:direct} 
\setcounter{equation}{0} 

The colorons in this class of models are identical to those introduced
in the flavor-universal coloron model of ref. \cite{coloron}.  As
discussed in \cite{colophen}, searches in dijet final states should be
the most powerful way of locating heavy colorons.  Searches in $b\bar b$
and $t\bar t$ offer no particular advantage in searching for the
flavor-universal colorons in our class of models.  This is in contrast
to the case of the topgluons of topcolor \cite{topcolor} and
topcolor-assisted technicolor \cite{topassist}.

As discussed earlier, constraints on the low-energy effective theory for
our class of models limit the value of coupling $\kappa_3$ to lie quite
close to the critical value $\approx 2$.  This means that the coloron
cannot be very light: if we estimate the minimum coloron mass for
$\kappa_3 = 2$ by requiring the coloron contribution (\ref{delrhoc}) to
$\Delta\rho^{\ast}$ to be less than $0.4\%$, we find $M_c \gae 1.6$ TeV;
including the $Z'$ contributions to $\Delta\rho^{\ast}$ would only
strengthen the bound.  A coloron of this large a mass lies above the
reach of published searches for new particles decaying to dijets
\cite{bump}.  Moreover, the large value of $\kappa_3$ implies that the
coloron's width
\beq
\Gamma_C \approx M_C \kappa_3 \bigg{[}{5 \over 6} + {1 \over
  6}\bigg{(}1-{{m_t}^2 \over M_{C}^2}\bigg{)}
\sqrt{1-{4m_t^2 \over M_C^2}} \bigg{]}
\eeq
is approximately twice its mass.  Future searches for narrow resonances
will not be appropriate for finding these colorons.  A more promising
approach would employ the strategies of compositeness searches, which
focus on high-$E_T$ enhancement of single-jet inclusive and dijet
spectra \cite{comp-spec} or alteration of the dijet angular
distributions \cite{comp-ang}.  At energies well below $M_C$, the
effects of coloron exchange on hadronic scattering are approximated by
those of the color-octet quark contact interaction (\ref{effcoleq}).  If
experiment set a limit $\Lambda_{octet} > X$ TeV on a color-octet
contact interaction
\beq 
- {g_o^2\over {2! \Lambda_{octet}^2}}    
\left(\bar{q} {\gamma}^{\mu} {{{\lambda}^{a}} \over2} q\right) 
\left(\bar{q} {\gamma}_{\mu} {{{\lambda}_{a}} \over2} q\right)   
\eeq  
with the usual convention $g_o^2/4\pi \equiv 1$, this would imply a
limit $M_C > \sqrt{2} X$ TeV for our class of models in which $\kappa_3
\approx 2$.

Existing limits on the mass of the $Z'$ boson are not very stringent.
For example, Tevatron bounds \cite{dylim} on new contributions to the
dilepton ($ee$ or $\mu\mu$) mass spectrum from interactions like
(\ref{effzpeq}) set no useful limit on our class of models because the
$Z'$ coupling to first generation fermions is so small.  The strongest
limits are derived in ref. \cite{ZZ} by considering the contributions to
electroweak observables of a $Z'$ like the one in our class of models
(called an ``optimal'' $Z'$ in \cite{ZZ}).  These calculations set a
95\% c.l.  lower bound of $290$ GeV on the $Z'$ for $\kappa_1 \approx
0.13$.  For other values of $\kappa_1$, the $Z'$ must be heavier; a $Z'$
mass less than a TeV is allowed for $.014 \lae \kappa_1 \lae .23$.
Including the effects of the colorons and q-pions on electroweak
observables would presumably strengthen the lower bounds on $M_{Z'}$, as
coloron exchange tends to increase $\Delta\rho^{\ast}$ (c.f.
\ref{delrhoc}) and the q-pions will contribute to hadronic $Z$
decays\footnote{Indeed, the presence of a full set of q-pions offers the
  possibility of new effects controlled by the scale $M_{q-pion}$ that
  may offset the large negative contributions to $R_b$ from top-pions
  and bottom-pions (and similar effects on $R_c$) found in
  \cite{gburddkom} for topcolor models.  This will be addressed in
  future work.}

Future experiments measuring production of third-generation fermions
($\tau^+\tau^-$, $\bar{b} b$, $\bar{t} t$) have the greatest potential
to find signs of the $Z'$ boson.  Consider, for example, looking for an
excess in $e^+ e^- \to \tau^+\tau^-$ in 50 fb$^{-1}$ of NLC data taken
at $\sqrt{s} = 500$ GeV.  Because the $Z'$ boson's decay width
\beq
\Gamma_{Z'}=M_{Z'}{\kappa_1 \over 3}\bigg{[}{20 \over 3}
\bigg{(}{\alpha_{Y} \over \kappa_1}\bigg{)}^2 + 
{23 \over 12} + {17 \over 12}\bigg{(}1-{{m_t}^2 \over M_
{Z'}^2}\bigg{)} \sqrt{1-{4m_t^2 \over M_{Z'}^2}}\bigg{]}
\eeq
is a large fraction of its mass (e.g., $\Gamma_{Z'} \approx .5 M_{Z'}$
for $\kappa_1 = .5$), we use the $s$-dependent width in the
cross-section; this renders our results insensitive to the exact value
of $\kappa_1$.  Assuming a 50\% efficiency for identifying tau pairs and
requiring a excess over the standard model prediction for $e^+e^- \to
[\gamma, Z] \to \tau^+ \tau^-$ of $(N^{\tau\tau} - N^{\tau\tau}_{SM})
\ge 5 \sqrt{N^{\tau\tau}_{SM}}$, the effects of a 2.7 TeV $Z'$ boson
with $\kappa_1 \le 1$ could be visible.  At a 1.5 TeV NLC with 200
fb$^{-1}$ of data, the reach in $M_{Z'}$ extends to 6.6 TeV.

\section{Conclusions} 
 
We have examined the low-energy effective theory and phenomenology of a
class of technicolor models with flavor-universal extended color
interactions and a generation-distinguishing extended hypercharge
sector.  Such models are found to be capable of dynamically producing a
top quark condensate that preferentially enhances the mass of the top
quark.  Moreover, flavor-changing neutral currents are less dangerous
here than in models where the color sector couples differently to the
third generation.  Constraints from $Z$-pole physics and $U(1)$
triviality single out the region of coupling-constant parameter space
where $\kappa_3 \approx 2$ and $\kappa_1 \lae 1$ for further study.
Electroweak physics presently constrains the $Z'$ boson in these models
to weigh at least 290 GeV, while the octet of flavor-universal colorons
must have a mass of at least 1.6 TeV.  Future studies of jet physics at
hadron colliders have the potential to uncover evidence of the colorons,
while data on pair-production of third-generation fermions at $e^+ e^-$
machines can help discover the $Z'$.
  
\vspace{12pt} \centerline{\bf Acknowledgments} \vspace{2mm} 
 
We thank R.S.~Chivukula, N.~Evans, C.T.~Hill, K.D.~Lane, and T.~ Rizzo
for useful discussions and comments on the manuscript.  E.H.S. is
grateful for the hospitality of the Aspen Center for Physics during the
inception of this work and that of the Theoretical Physics Group at
Fermilab during its completion.  E.H.S. acknowledges the support of the
Faculty Early Career Development (CAREER) program and the DOE
Outstanding Junior Investigator program.  {\em This work was supported
  in part by the National Science Foundation under grant PHY-9501249,
  and by the Department of Energy under grant DE-FG02-91ER40676.}

\newcommand{\np}{{\it Nucl.\ Phys.}\ {\bf B}} 
\newcommand{\pr}{{\it Phys.\ Rev.}\ } 
\newcommand{\prd}{{\it Phys.\ Rev.}\ {\bf D}} 
\newcommand{\prp}{{\it Phys.\ Rep.}\ } 
\newcommand{\prl}{{\it Phys.\ Rev.\ Lett.}\ } 
\newcommand{\pl}{{\it Phys.\ Lett.}\ {\bf B}} 
\newcommand{\ptp}{{\it Prog.\ Theor.\ Phys.}\ } 
\newcommand{\ap}{{\it Ann.\ Phys.}\ } 
\newcommand{\intl}{{\it Int.\ J.\ Mod.\ Phys.}\ {\bf A}} 
\newcommand{\mpl}{{\it Mod.\ Phys.\ Lett.}\ {\bf A}}

\end{document}